\documentclass{article}
\usepackage{amsmath,latexsym}
\numberwithin{equation}{section}
\newcommand{\norm}[1]{\lVert#1\rVert}
\newcommand{\abs}[1]{\lvert#1\rvert}
\allowdisplaybreaks

\begin{document}

\begin{center}
\title{\bf
${\boldsymbol{B\to D^{\ast\ast}}}$
semileptonic decay in covariant quark models \`a la Bakamjian Thomas}
\author{V. Mor\'enas$^a$ \\[5mm] 
 A. Le Yaouanc, L. Oliver, O. 
P\`ene and J.-C. Raynal$^b$}\par
\maketitle
{$^a$ Laboratoire de Physique Corpusculaire\\
Universit\'e Blaise Pascal - CNRS/IN2P3
F-63177 Aubi\`ere Cedex, France\\
$^b$ Laboratoire de Physique Th\'eorique et Hautes
Energies\footnote{Laboratoire
associ\'e au
Centre National de la Recherche Scientifique - URA D00063
\\e-mail: morenas@clrcls.in2p3.fr, pene@qcd.th.u-psud.fr}}\\
{Universit\'e de Paris XI, B\^atiment 211, 91405 Orsay Cedex,
France}
\end{center}

\begin{abstract}

Once chosen the dynamics in one frame, for example the rest frame,
the Bakamjian and Thomas method allows to define
 relativistic quark models in any frame. These models
 have been shown to provide, 
in the heavy quark limit, fully covariant current form factors as
matrix elements of the quark current operator. They also verify the 
Isgur-Wise scaling and give a slope parameter $\rho^2>3/4$ for all
the possible choices of the dynamics.
 In this paper we study the $L=1$ excited states and derive
 the general formula, valid for any dynamics,
 for the scaling invariant form factors 
$\tau_{1/2}^{(n)}(w)$ and $\tau_{3/2}^{(n)}(w)$. We also 
check the Bjorken-Isgur-Wise sum rule already demonstrated 
elsewhere in this class of models.

\end{abstract}
\begin{flushright} LPTHE Orsay-96/12\\PCCF RI 9601 \\ hep-ph/9605206
\end{flushright}
\newpage

\section{Introduction}

It was recently noticed \cite{raynal} that it is possible, in the 
heavy mass limit, to formulate fully covariant quark models for form 
factors in which the current acts in the standard way on the heavy 
quark while the other quarks remain spectators. Following Bakamjian and Thomas
(BT) \cite{BT}, given the wave 
function in, say, the rest frame\footnote{Any starting frame can be 
chosen, for example the infinite momentum frame although some care is 
then needed.}, the hadron wave functions are defined in any frame through a
unitary transformation, in such a way that 
Poincar\'e algebra is satisfied. It was shown in \cite{raynal} that 
the $\rho^2$ Isgur-Wise slope parameter was bounded in this class of 
model: $\rho^2>0.75$. 

It is known that quark models are of special value when excited 
states are considered, since then no other hadronic method is 
available: lattice simulation as well as QCD sum rules are 
practically restricted to ground states because they use euclidean 
continuation.
It is therefore tempting to apply this covariant approach to the 
$B\to D^{\ast\ast}$ decays. In \cite{alain} it has been shown that 
these covariant quark models satisfy the Bjorken sum rule
 \cite{bjorken}, \cite{IW}. This is 
not a trivial achievement. It comes in this class of models because 
the boost of the wave functions is a unitary transformation that 
keeps the closure property of the Hilbert space in all frames.  

In \cite{alain} the precise formulae for the $\tau_{1/2}^{(n)}(w)$ and
$\tau_{3/2}^{(n)}(w)$ have not been computed except for $w=1$.
In view of practical phenomenology of the $B\to D^{\ast\ast}$ decays,
$\tau_{1/2}^{(n)}(w)$ and
$\tau_{3/2}^{(n)}(w)$ are obviously needed for any $w$. This is our main goal
in the
present paper. We do not want here to make a choice for our preferred internal
wave function. These formulae may be used by anyone who wants to apply the BT
method in the heavy mass limit to its preferred set of rest-frame, or
infinite momentum frame or whichever frame wave functions. The Ansatz
\cite{close}, although not explicitly of the BT type and computed in
particular frames, would probably be obtained by the BT method with the
harmonic oscillators. The same comment applies to the calculation of
$B\to D^{\ast\ast}$ \cite{wambach}. In deriving the formulae for
$\tau_{1/2}^{(n)}(w)$ and $\tau_{3/2}^{(n)}(w)$, we will also try to repeat,
as much in a transparent way as possible, the content of the BT method
applied to the heavy quark limit. This will be done in section \ref{sec-cov}.
In section  \ref{sec-tau} we will compute the $\tau_{1/2}^{(n)}(w)$ and
$\tau_{3/2}^{(n)}(w)$. In section \ref{sec-bj} we will derive directly from
the latter formulae the Bjorken-Isgur-Wise sum rule.

\section{Covariant quark models of form factors in the heavy mass 
limit}\label{sec-cov}

\subsection{Framework}
The main purpose of this model is to provide a way to implement 
covariance
in the calculation of form factors. These form factors appear through 
matrix elements
relating the initial and final states of the hadrons. We assume here 
a spectator quark
model, that is, of all the quarks building the hadron, only one 
(labeled as $1$) is the
active particle. With this hypothesis, the amplitude of any process 
can be written as:
\begin{equation*}
\langle\Psi'| O |\Psi\rangle  =  \int \frac{d{\vec 
p}^{\,\prime}_1}{{(2\pi)}^3}
\frac{d\vec p_1}{{(2\pi)}^3}\,  \frac{d\vec p_2}{{(2\pi)}^3} 
\sum_{s'_1,s_1,s_2}
\Psi_{s'_1,s_2}({\vec p}^{\,\prime}_1, \vec p_2)^{\displaystyle *} 
O({\vec p}^{\,\prime}_1,\vec p_1)_{s'_1,s_1}  \Psi_{s_1,s_2}(\vec 
p_1, \vec p_2)
\end{equation*}\noindent
where we have used the so-called one particle variables, momentum 
$\vec p_i$ and
spin $\vec S_i$, and where $O({\vec p}^{\,\prime}_1,\vec 
p_1)_{s'_1,s_1}$ is the matrix
element of the {\em free} one-particle operator $O$ between one 
particle states of the form
$| \vec p, s \rangle$
(we only take into account two quarks since we are dealing with mesons; 
the generalization to $n$ particles is straightforward \cite{raynal}).
\par\vspace{5mm}
Now, the covariance is introduced by expressing the relativistic 
2-particle bound states
$\Psi_{s_1,s_2}(\vec p_1, \vec p_2)$ as a representation of the full
Poincar\'e group, following the Bakamjian-Thomas formalism. It turns 
out that this process is
made easier if we change the variables that characterize the state 
$\Psi$, that is, if
we introduce the total momentum $\vec P$, $(\vec P=\vec p_1+\vec 
p_2)$,
the internal momenta $\vec k_1, \vec k_2$, $(\vec k_1+\vec 
k_2=\vec 0)$, and the internal spins $\vec S'_1, \vec S'_2$,
leading to another wave function $\Psi^{int}_{s_1, s_2}
(\vec P,\vec k_2)$ which is related to the previous one by the {\em 
unitary} transformation:
\begin{equation}
\Psi_{s_1, s_2}(\vec p_1,\vec p_2) \ =\ 
\sqrt{\frac{p_1^o+p_2^o}{M_o}}\,
\frac{\sqrt{k^o_1\,k^o_2}}{\sqrt{p^o_1\,p^o_2}}\,\sum_{s'_1, s'_2}
(D({\mbox{\boldmath $R$}}_1)_{s_1,s'_1}\,D({\mbox{\boldmath 
$R$}}_2)_{s_2,s'_2})\;
\Psi^{int}_{s'_1, s'_2}(\vec P, \vec k_2)\label{uni}
\end{equation}
Let us explain the notation used in this last equation:
\begin{itemize}
\item {$p_i^o$}$\ =\ \sqrt{\vec p_i^{\,2}\,+\,m_i^2}$
\item {$M_o$}$\ =\ \sqrt{(\Sigma p_j)^2}$
\item {$k_2$}$\ =\ \mbox{\boldmath $B$}^{-1}_{\Sigma p_j} p_2$
\item {$D({\mbox{\boldmath $R$}}_i)_{s_i,s'_i}$} is the 
$(s_i,s'_i)$ matrix element for a spin $1/2$ particle of the Wigner
rotation $\mbox{\boldmath $R$}_i$, which describes the 
connection spin$\leftrightarrow$internal spin and which is given by:
\begin{equation}
\mbox{\boldmath $R$}_i = \mbox{\boldmath
$B$}^{-1}_{p_i} \mbox{\boldmath $B$}_{\Sigma p_j} \mbox{\boldmath 
$B$}_{k_i}\label{c}
\end{equation}
where $\mbox{\boldmath $B$}_{p}$ is the Lorentz boost $(\sqrt{p^2}, 
\vec 0) \to p$.\end{itemize}\par\noindent
Note that there is no apparent dependence upon $\vec k_1$ owing to
the relation which defines the internal momenta.\par\vspace{5mm}
Returning now to the Poincar\'e group transformation laws, their 
generators are defined according to:
\begin{equation}\label{d}
\left\{  \ {
\begin{split}
\mbox{space translations}\ &:\ \vec P\\ 
\mbox{time translations}\ &:\ {\displaystyle H\ =\ P^o\ =\ \sqrt{{\vec P}^2\,+
\,M^2}}\\ 
\mbox{rotations}\ &:\ {\displaystyle \vec J\ =\ -i\vec P\times\frac{\partial}
{\partial 
\vec P}\ +\ \vec  S}\\ 
\mbox{boosts}\ &:\ {\displaystyle \vec K\ =\ -\frac{i}{2}\left\{  {P^o\, 
,\,\frac{\partial}{\partial \vec P}}
\right\}\ -\ \frac{\vec P \times\vec S}{P^o\,+\,M}}
\end{split} }\right.
\end{equation}
\begin{equation*}
\mbox{where}\hspace{30mm}  \vec S\ =\ (\vec S'_1\,+\,\vec S'_2)\ -\ 
i\vec k_2 \times
\frac{\partial}{\partial\vec k_2}\hspace{10mm}\mbox{is the spin of 
the meson}
\hspace{25mm}
\end{equation*}
\noindent  These generators depend on the
interaction 
through the mass operator $M$. Any given quark model corresponds to one choice
of the mass operator $M$, in order, for example, to describe the mass spectrum
at best. In this paper we do not want to enter into the discussion of the
best choice for the latter operator. We simply need $M$ to depend only on the
internal variables and to be invariant by
rotation in order to obey Poincar\'e algebra.
The mass $M_0=\sqrt{(\Sigma p_j)^2}$ introduced above is not the real mass. It
is just a tool to define properly the unitary change of variables (\ref{uni}).
Only in the limit when the quark interaction would vanish, would $M_0$ become
equal to $M$. In such a limit, the generators in (\ref{d}) would be the
standard Poincar\'e generators for the considered set of free particles. As
such it would naturally obey Poincar\'e algebra. In some sense, the trick used
above in order to have the correct Poincar\'e algebra even when interaction is
present is to mimic, so to say, the behavior of free particles by the
definition of $p_i^0$, $M_0$. This does not mean that we make a weak
interaction approximation.
One may say that the change of variables (\ref{uni}) is inspired from the
non-interacting case, but it may be applied whatever the strength of the
interaction is. 
\par\vspace{5mm}
It is then possible to construct \cite{raynal} the wave function of the 
bound state of quarks moving with total momentum $\vec P$ and we get:
\begin{equation}
\Psi^{int}_{s_1,s_2}(\vec P, \vec k_2)=  (2\pi)^3\,
\delta(\vec p_1+\vec p_2 - \vec P)\;  \varphi_{s_1,s_2}(\vec k_2)
\end{equation}
where $\varphi_{s_1,s_2}(\vec k_2)$ is an eigenstate of $M$, $\vec S$ and $S_z$,
so that the relativistic wave function expressed with the momenta 
$\vec p_i$ and spins $\vec S_i$ reads:
\begin{equation*}
\Psi_{s_1, s_2}(\vec p_1,\vec p_2)=
\sqrt{\frac{p^o_1+p^o_2}{M_o}}\frac{\sqrt{k^o_1\,k^o_2}}{\sqrt{p^o_1\,p^o_2}}
\;\sum_{s'_1, s'_2}D({\mbox{\boldmath $R$}}_1)_{s_1,s'_1}\,
D({\mbox{\boldmath $R$}}_2)_{s_2,s'_2}\;(2\pi)^3\,
\delta(\vec p_1+\vec p_2 - \vec P)\;\varphi_{s'_1,s'_2}(\vec k_2)
\end{equation*}
\noindent Notice that for $\vec P=\vec 0$, this last formula gives
\begin{equation*}
\Psi^{(\vec P=\vec 0)}_{s_1,s_2}(\vec p_1,\vec p_2)\ =\ (2\pi)^3\, 
\delta(\vec p_1+\vec p_2)\; 
\varphi_{s_1,s_2}(\vec p_2)
\end{equation*}\noindent
which just expresses the fact that $\varphi$ is the rest-frame 
internal wave function.\par
\vspace{5mm}
Putting all these things together, we get the new following workable 
relation:
\begin{multline}\label{h}
\langle\vec P'| O |\vec P\rangle\ =\ \int \frac{d\vec 
p_2}{{(2\pi)}^3}\, 
\sqrt{\frac{(p^{\prime o}_1+p^o_2)(p^o_1+p^o_2)}{M'_oM_o}}\,
\frac{\sqrt{k^{\prime o}_1k^o_1}}{\sqrt{p^{\prime o}_1p^o_1}}
\frac{\sqrt{k^{\prime o}_2k^o_2}}{\sqrt{p^o_2p^o_2}} \\
\sum_{s'_1,s'_2}\,\sum_{s_1,s_2}   
{\varphi'_{s'_1,s'_2}(\vec k'_2)^{\displaystyle*}}
[D(\mbox{\boldmath $R$}_1^{\prime-1})O(\vec p^{\,\prime}_1,\vec 
p_1)D(\mbox{\boldmath $R$}_1)]_{s'_1,s_1}\,
D(\mbox{\boldmath $R$}_2^{\prime-1}\mbox{\boldmath 
$R$}_2)_{s'_2,s_2}\,\varphi_{s_1,s_2}
(\vec k_2)
\end{multline}\noindent
since $\vec {p_2}^{\prime}=\vec p_2$ for the spectator quark. 
At this stage, this matrix element is not covariant in general, 
because the current
operator $O$ is not covariant with respect to the transformations 
\eqref{d} which contain the
interaction. However, in the limit where
the masses $m_1$ and $m_1^{\prime}$ of the quark $1$ tend to 
infinity, this last
equation becomes fully covariant as shown in \cite{raynal}.\par\vspace{5mm}
So, up to now, we have replaced the knowing of a relativistic wave 
function of a moving
bound state by the knowing of a rest-frame wave function, which is an 
eigenstate of a
mass operator (in this case the hamiltonian) which does not have to 
be relativistic.
Of course, the physics lies in the complete determination of those 
rest-functions
$\varphi_{s_1,s_2}(\vec k_2)$,
and in the expression of the current operator.\par
\noindent Another point which must be emphasized is the unitary 
nature of the transformations
which have been used.
Unitarity ensures that, in any frame, all sets of wave functions are 
orthonormal and satisfy the
closure relation. This will prove to be essential for the derivation of the 
Bjorken 
sum rule (see section \ref{sec-bj}).\par\vspace{5mm}
From now on, we will assume that the quark $1$ has an infinite mass 
to enable the Heavy Quark
Symmetries (flavor and spin symmetries) and the covariance of the 
matrix elements.
Besides, we will denote the internal spins $\vec s_i$  and no longer 
$\vec S'_i$.

\subsection{The wave functions}
In the rest frame $(\vec P=\vec 0)$, the generator of rotations 
writes:
\begin{equation}\label{Jj}
\vec J\ =\ \vec s_1\ +\ \vec j.
\end{equation}

In quark models, $\vec j$ equals $\vec s_2\,+\,\vec l$ and $\vec 
l=\displaystyle{-i\vec k_2\times
{\partial}/{\partial\vec k_2}}$ is the relative orbital angular 
momentum of the second
quark, which is equal to $1$ in the case of $D^{**}$ mesons. However, beyond
quark models, the decomposition \eqref{Jj} is very useful in general in the
heavy quark limit. Indeed, according
to Heavy Quark Symmetry, the spin $\vec s_1$ of the heavy quark is conserved,
i.e. it commutes with the Hamiltonian. It then results from conservation of
the total angular momentum $\vec J$ that the angular momentum $\vec j$ is
also conserved. It is then natural
to label the lightest parity-even states ($l=1, P$-wave in the quark models),
according to the values of $j$, namely $j=1/2$ and
$j=3/2$. It is known \cite{IW} that all hadrons within $j=1/2$, respectively
$j=3/2$, multiplet, are related by the heavy quark symmetry\footnote{All these
properties are valid in the model, and we shall use them in the next
subsection.}. Finally, when
combined with the heavy quark spin $s_1$, we  get two distinct multiplets: 
one with $j=1/2$
and $J^P=0^+\ \mbox{or}\ 1^+$, and the other with $j=3/2$ and 
$J^P=1^+\ \mbox{or}\ 2^+$.
The corresponding states will be denoted respectively, using the notation
$|j J^P>$:
\begin{equation*}
|{\scriptstyle\frac{1}{2}}\,0^+\rangle\hspace{15mm}|{\scriptstyle\frac{1}{2}}
\,1^+\rangle
\hspace{15mm}
|{\scriptstyle\frac{3}{2}}\,1^+\rangle\hspace{15mm}|{\scriptstyle\frac{3}{2}}\,
2^+\rangle \nonumber
\end{equation*}

\subsubsection{Generic form of the ${\boldsymbol \varphi}$'s}
How are we going to write the rest-frame internal wave functions 
? Recall that they are eigenstates of the mass operator $M$. In the model,
this operator is assumed to
be rotationally invariant, to depend only upon the internal variables 
and, of course,
to conserve parity. So it commutes with $\vec J$, which in turn 
commutes with the hamiltonian
$H$. Now, the heavy quark spin symmetry implies the invariance with 
respect to $\vec s_1$ of the
system, that is $[H,\vec s_1]\equiv [M,\vec s_1]=0$. As a consequence, 
$[H,\vec j]\equiv [M,\vec j]=0$
and the total Hilbert space $\cal H$ of the $\varphi$'s can be
factorized as a tensorial product of a spin space ${\cal H}_{\vec 
s_1}$, related to the spin
$\vec s_1$, and of a ``spin-orbit'' space ${\cal H}_{\vec j}$ which 
describes the internal
spin $\vec s_2$ of the light quark and the relative orbital angular 
momentum $\vec l$.
We must therefore build a representation of $\vec j =\vec s_2+\vec 
l$. But we know a priori
nothing about $[H,\vec l]$ or $[H,\vec s_2]$ so we cannot reproduce 
the same kind of argument
used for the decomposition of $\cal H$ into ${\cal H}_{\vec 
s_1}\otimes{\cal H}_{\vec j}$.
Yet, we are concerned by the two lightest parity even multiplets ($j=1/2$ and
$j=3/2$), 
with parity $+1$. They all correspond to the same value of the
orbital momentum,
$l=1$. Indeed, due to parity conservation, $l=1$ cannot be mixed with $l=0,2$,
and due to the conservation of $\vec j$, it cannot be mixed with $l=3,5$
which cannot produce $j=1/2, 3/2$ when combined with $s_2=1/2$.
Whence, the only way of constructing
this representation of
$\vec j$ is $\vec l\otimes \vec s_2$ with $l=1$.

However, we must stress again that the radial part of the wave functions and
the energies depend on $j$. The four $l=1, s_2=1/2$ states are not degenerate
in energy due to an $\vec l\cdot \vec s_2$ force which does not vanish in the
heavy quark limit. Since the eigenvalue of $\vec l\cdot \vec s_2$ is equal to
$(j(j+1)-l(l+1)-s_2(s_2+1))/2=j(j+1)/2-1-3/8$, we see that the four states
will split as expected into two multiplets corresponding to $j=1/2$ and
$j=3/2$ respectively. The spin and orbital parts of the wave functions
combine through Clebsch-Gordan coefficients to build up the $j=1/2, 3/2$
eigenstates. As a consequence the $^3P_1$ and the $^1P_1$ states will mix to
produce the $j$ eigenstates.
\par\noindent
Finally, we can factorize the wave function:
\begin{equation}
\Psi^{j,J^P}_{M,s_1,s_2}(\vec p)\ =\ \left( {jJ^P}_{M,s_1,s_2}\right)(\hat p)\,
\sqrt{\frac{4\pi}{3}}\;\norm{\vec p}\;\phi_j(\norm{\vec p}^2)\label{2.6}
\end{equation}
\noindent where
\begin{equation}
\left( {jJ^P}_{M,s_1,s_2}\right)(\hat p)\ =\ \sum_{m}\langle\,j\ m\  
{1/2}\ M-m\ |\ J\ M\,\rangle\,\chi^{M-m}_{s_1}\,\sum_{m'}\langle\,1\ m'\ 
{1/2}\ m-m'\ |\ j\ m\,\rangle\,\chi^{m-m'}_{s_2}\,
\;Y^{m'}_1(\hat p)
\end{equation}
$Y^m_l(\hat p)$ being the spherical harmonics $\langle 
\theta, \phi|lm\rangle$
and $\phi_j(\norm{\vec p}^2)$ a radial function which remains to be determined,
and is normalized according to
\begin{equation*}
\int \frac{d\vec p}{(2\pi)^3} \frac {p^2}3 \vert\phi_j(\norm{\vec p}^2)
\vert^2=1. 
\end{equation*}
The
unusual writing of the radial part in \eqref{2.6} is a matter of convenience,
meant to simplify further calculation. Moreover,
$\chi^m_s$ is the column matrix defined by:
\begin{equation*}
\chi^{+1/2}\ =\ \begin{pmatrix}
1\cr 0\end{pmatrix}\hspace{25mm}\chi^{-1/2}\ =\ \begin{pmatrix}0\cr 
1\end{pmatrix}
\end{equation*}
\noindent From now on, we will often identify the indices $1/2$ with
$1$ and $-1/2$ with $2$
as a way of spotting the matrix elements which are related to spin.
Consequently, the $\chi^m_s$ equal $\delta_{m\,s}$.\par\vspace{5mm}

Let us concentrate for awhile on the spin-orbital part of the wave function.
The calculations are made easier if we deal with another composition
of $\vec l$, $\vec s_1$ and $\vec s_2$ in order to get $\vec J$, that
is if, instead of writing $\vec s_1\otimes (\vec s_2\otimes \vec l)$, we
use $(\vec s_1\otimes \vec s_2)\otimes \vec l$. With that
decomposition, we obtain the so-called $|^{2S+1}P_J\rangle$ states,
which can be related to the $|jJ^P\rangle$ through the ``$6j$''
coefficients according to:
\begin{equation*}
|jJ^P\rangle\ =\ \sum_S\,(-)^{s_1+s_2+l+J}\,\sqrt{(2S+1)(2j+1)}\,
\left\{\begin{matrix}
s_1&s_2&S\\ l&J&j\\
\end{matrix}\right\}\,|^{2S+1}P_J\rangle
\end{equation*}
which reads in the present case
\begin{equation*}
\left\{ \ {
\begin{split}
{\displaystyle |{\scriptstyle\frac{1}{2}}\,0^+\rangle}\ &=\ 
{\displaystyle |^{3}P_0\rangle}\\ 
{\displaystyle |{\scriptstyle\frac{1}{2}}\,1^+\rangle}\ &=\ 
{\displaystyle -\frac{1}{\sqrt 3}\,|^{1}P_1\rangle\ +\ 
\sqrt{\frac{2}{3}}\,|^{3}P_1\rangle}\\ 
{\displaystyle |{\scriptstyle\frac{3}{2}}\,1^+\rangle}\ &=\ 
{\displaystyle \sqrt{\frac{2}{3}}\,|^{1}P_1\rangle\ +\ 
{\frac{1}{\sqrt 3}}\,|^{3}P_1\rangle}\\ 
{\displaystyle |{\scriptstyle\frac{3}{2}}\,2^+\rangle}\ &=\ 
{\displaystyle |^{3}P_2\rangle}
\end{split} }\right.
\end{equation*}
with
\begin{multline*}\begin{aligned}
\left( {^{2S+1}P_J}_{s_1,s_2,M}\right)(\hat p)&=\sum_m
\langle\,1\ m\ S\ M\!-\!m\ |\ J\ M\,\rangle \,Y_1^m(\hat p)\,
\sum_{a,b}\,\langle\,{1/2}
\ a\ {1/2}\ b\ |\ S\ M\!-\!m\,\rangle\,
(\chi^a)_{s_1}\,(\chi^b)_{s_2}\\
&=\sum_m
\langle\,1\ m\ S\ M\!-\!m\ |\ J\ M\,\rangle \,
\langle\,{1/2}
\ s_1\ {1/2}\ s_2\ |\ S\ M\!-\!m\,\rangle\,
Y_1^m(\hat p)\end{aligned}\end{multline*}
$$\text{since}\ (\chi^m)_{s}=\delta_{m\,s}$$

It may also be useful
to remark that the Clebsch-Gordan coefficient $\langle\,{1/2}
\ s_1\ {1/2}\ s_2\ |\ S\ m\,\rangle$ can be written as the matrix
element of a combination of Pauli matrices according to:
\begin{equation}\label{g}
\begin{split}
\langle\,{1/2}\ s_1\ {1/2}\ s_2\ |\ 0\ m\,\rangle&=
\frac{i}{\sqrt 2}\,\left( \sigma_2\right)_{s_1\,s_2}\\
\langle\,{1/2}\ s_1\ {1/2}\ s_2\ |\ 1\ m\,\rangle&=
\frac{i}{\sqrt 2}\,\left[\left({\vec e}^{\,(m)}\cdot{\vec \sigma}\right)
\sigma_2\right]_{s_1\,s_2}
\end{split}
\end{equation}
where we have used the standard basis:
\begin{equation*}
\left\{ \ {
\begin{split}
{\displaystyle \vec e^{\,(+1)}}\ &=\ {\displaystyle -{\frac{1}{\sqrt 2}}
\left( \vec e_x + i\,\vec e_y\right)}\\
{\displaystyle \vec e^{\,(0)}}\ &=\ {\displaystyle \vec e_z}\\
{\displaystyle \vec e^{\,(-1)}}\ &=\ {\displaystyle \frac{1}{\sqrt 2}
\left( \vec e_x - i\,\vec e_y\right)}
\end{split} }\right.
\end{equation*}
(The demonstration of these relations can be carried out by evaluating
the right hand sides of \eqref{g} and then by comparing
the results with the values of the Clebsch-Gordan coefficients).
\par\vspace{5mm}
Finally, by substituting the expressions of the $Y_l^m(\hat p)$ in spherical
coordinates, the
$\left| {^{2S+1}P_J}\right>$ spin-orbital wave functions write:
\begin{equation*}
\left\{ \ {
\begin{split}
{\displaystyle \left( {^{3}P_0}_{s_1,s_2}\right)(\hat p)}\ &=\ 
{\displaystyle -\frac{i}{\sqrt{8\pi}}\,
\left[ {\left( {{\vec \sigma} \cdot {\vec p}}\right)\sigma_2}
\right]_{s_1\,s_2}\cdot\dfrac{1}{\norm{\vec p}}
}\\
{\displaystyle \left( {^{1}P_1}_{s_1,s_2,m}\right)(\hat p)}\ &=\ 
{\displaystyle i\sqrt{\frac{3}{8\pi}}\,
\left[ {\left( {{\vec e^{\,(m)}} \cdot {\vec p}}\right)\sigma_2}\right]_{s_1\,
s_2}\cdot\dfrac{1}{\norm{\vec p}}
}\\
{\displaystyle \left( {^{3}P_1}_{s_1,s_2,m}\right)(\hat p)}\ &=\ 
{\displaystyle \sqrt{\frac{3}{16\pi}}\,
\left[ {\left( {{\vec e^{\,(m)}} \cdot (\vec p \wedge \vec \sigma)}\right)
\sigma_2}\right]_{s_1\,s_2}
\cdot\dfrac{1}{\norm{\vec p}}
}\\
{\displaystyle \left( {^{3}P_2}_{s_1,s_2,m}\right)(\hat p)}\ &=\ 
{\displaystyle -i\sqrt{\frac{3}{8\pi}}\,
\left[ {\left( {\sigma^ip^j\,e^{\,(m)}_{ij}}\right)\sigma_2}\right]_{s_1\,s_2}
\cdot\dfrac{1}{\norm{\vec p}}} 
\end{split} }
\right.
\end{equation*}
where $e^m_{ij}$ is the rest-frame polarization tensor,
that is a symmetrical tensor with vanishing spur.\par\vspace{5mm}
We are now able to go one step further into the determination of the
transition amplitude
$\langle\vec P'| O |\vec P\rangle$. In the present case, the state
$|\vec P\rangle$ represents
a pseudoscalar state $^1S_0$ and the state $|\vec P'\rangle$ one of the
$\left( {^{2S+1}P_J}\right)_j$ states described above.
In other words, the $\varphi_{s_1,s_2}(\vec k)$'s take the form:
\begin{eqnarray*}
\varphi_{s_1,s_2}(\vec k_2)&=&\frac{i}{\sqrt 2}\,(\sigma_2)_{s_1\,s_2}\,
\varphi(\norm{\vec k_2}^2)\\
\varphi'_{s'_1,s'_2}(\vec k'_2)&=&\frac{i}{\sqrt 2}\,
(\chi\sigma_2)_{s'_1\,s'_2}\,
\phi_j(\norm{\vec k'_2}^2)
\end{eqnarray*}
where
\begin {equation*}
\chi_{_{\left( {^{2S+1}P_J}\right)}}\ =\ 
\begin{cases}\vspace{3mm}
{\displaystyle \ -\frac{1}{\sqrt 3}\,(\vec \sigma\cdot\vec k'_2)}&
\text{for the } ^3P_0 \text{ state}\\
\vspace{3mm}
{\displaystyle \ (\vec e^{\,(m)}\cdot\vec k'_2)}& \text{for the } ^1P_1
\text{ state}\\ \vspace{3mm}
{\displaystyle \ -\frac{i}{\sqrt 2}\,\left[\vec e^{\,(m)}\cdot
(\vec k'_2\wedge\vec \sigma)\right]}
& \text{for the } ^3P_1 \text{ state}\\ \vspace{3mm}
{\displaystyle \ -\,({\sigma ^i} {k'_2}^j \,e^{\,(m)}_{ij})}&
\text{for the } ^3P_2 \text{ state}
\end{cases}
\end{equation*}
Then, by substituting these expressions into \eqref{h} and using the relation
$\sigma_2D(\mbox{\boldmath $R$})\sigma_2  =  D(\mbox{\boldmath $R$}^{-1})^t$,
we get:
\begin{multline}\label{em}
_j\langle{^{2S+1}P_J}| O |{^1S_0}\rangle\ =\ \int \frac{d\vec 
p_2}{{(2\pi)}^3}\, 
\sqrt{\frac{(p^{\prime o}_1+p^o_2)(p^o_1+p^o_2)}{M'_oM_o}}\,
\frac{\sqrt{k^{\prime o}_1k^o_1}}{\sqrt{p^{\prime o}_1p^o_1}}
\frac{\sqrt{k^{\prime o}_2k^o_2}}{\sqrt{p^o_2p^o_2}}\\
\times{\displaystyle \frac{1}{2}}\,\mathrm { Tr}\left[\chi^\dagger\,
D({\mbox{\boldmath $R$}'}_1^{-1})\,O(\vec
p^{\,\prime}_1,\vec p_1)\,D(\mbox{\boldmath $R$}_1)\,
D(\mbox{\boldmath $R$}_2^{-1}\mbox{\boldmath $R$}'_2)\right]
\;\phi_j(\norm{\vec k'_2}^2)^{\displaystyle*} \varphi(\norm{\vec k_2}^2)
\end{multline}

\subsection{Switching to Dirac notation}

As already stated, it can be shown directly on (\ref{h}) that the current
matrix element is covariant in the heavy mass
limit $m_1\to\infty$. We shall demonstrate covariance by changing to Dirac
notation, in which the current matrix elements
will eventually appear as {\it manifestly covariant.} This notation will also
be very useful to handle later
calculations.\par
\noindent The idea is to
insert the $2\times 2$ matrices, that appeared in the precedent sections,
into the $2\times 2$ upper left block of a $4\times 4$ matrix which is
then completed with zeros:
\begin{math}
\bigl( \begin{smallmatrix}
\chi&0\\ 0&0
\end{smallmatrix} \bigr)
\end{math}. This leads to the following formula:
\begin{multline}\label{a}
_j\langle{^{2S+1}P_J}| O |{^1S_0}\rangle=\int\,\frac{d\vec p_2}
{(2\pi)^3}\,\frac{1}{p^o_2}\,\frac{\sqrt{{u'}_o\,u_o}}{p^o_1\,{p'}^o_1}\,
\frac{k^o_1}{\sqrt{k^o_1+m_1}}\,\frac{k^o_2}{\sqrt{k^o_2+m_2}}\,
\frac{{k'}^o_1}{\sqrt{{k'}^o_1+m'_1}}\,\frac{{k'}^o_2}{\sqrt{{k'}^o_2+m_2}}\\
\times \frac{1}{16}\,\mathrm { Tr}\left\{ {
O\,(m_1+\rlap/p_1)\,(1+\rlap/u)\,(m_2+\rlap/p_2)\,(1+\rlap/u')\,\left( {
\mbox{\boldmath $B$}_{u'}{\chi '}^{\dagger}\mbox{\boldmath $B$}^{-1}_{u'}
}\right)\,(m'_1+\rlap/p'_1)
}\right\}\,\phi_j(\norm{\vec k'_2}^2)^{\displaystyle*}\varphi(\norm{\vec k_2}^2)
\end{multline}
where $u$ and $u'$ are defined by
\begin{equation*}
M_o\,u\ =\ p_1\,+\,p_2\ \ (u^2=1) \hspace{25mm}M'_o\,u'\ =\ p'_1\,+\,p_2 \ \ 
{(u'}^2=1)
\end{equation*}
The derivation of \eqref{a} is quite straightforward: it stems from
\begin{itemize}
\item the expressions of the Wigner rotation matrices as a product of three
boosts, see equation \eqref{c}
\item the
expression of the matrix $O(\vec p^{\,\prime},\vec p)$ as \cite{raynal}
\begin{equation*}
O(\vec p^{\,\prime}_1,\vec p_1)  = \frac{\sqrt{{m'}_1m_1}}{\sqrt{{p'}^0_1p^0_1}
}\;
\frac{1+\gamma^0}2 \mbox{\boldmath $B$}_{{p'}_1}^{-1}O
\mbox{\boldmath $B$}_{p_1} \frac{1+\gamma^0}2
\end{equation*}
In this last relation, the boosts $\mbox{\boldmath $B$}_p$ are written in the
Dirac representation
according to
\begin{equation*}
\mbox{\boldmath $B$}_p  =   \frac{m + \rlap/p\gamma^0}{\sqrt{2m(p^0+m)}}
\end{equation*}
\item the properties of $\mbox{\boldmath $B$}_{u}$, that is, for instance:
\begin{gather*}
\mbox{\boldmath $B$}_{u}(1,\vec 0)\ =\ u\hspace{25mm}(\text{idem for }
\mbox{\boldmath $B$}_{u'})\\
p_2\ =\ \mbox{\boldmath $B$}_{u}\,k_2\ =\ \mbox{\boldmath $B$}_{u'}\,{k'}_2
\end{gather*}
\item
the following formula,
which will also be used throughout the rest of this paper in order to
evaluate the terms $
\mbox{\boldmath $B$}_{u'}{\chi}^{\dagger}\mbox{\boldmath $B$}^{-1}_{u'}
$:
\begin{equation}
\mbox{\boldmath $B$}_{u}(\gamma_{\mu}\,x^{\mu})\mbox{\boldmath $B$}^{-1}_{u}
\ =\ \gamma_{\mu}\left( {\mbox{\boldmath $B$}_{u}x
}\right) ^{\mu}\label{b}
\end{equation}
(the $\gamma^{\mu}$ are forming a 4-vector), and which gives for example:
\begin{equation*}
\mbox{\boldmath $B$}_{u}\,\gamma_o\,\mbox{\boldmath $B$}^{-1}_{u}\ =\ 
\mbox{\boldmath $B$}_{u}\,(1,\vec 0)\gamma^{\mu}\,
\mbox{\boldmath $B$}^{-1}_{u}\ =\ 
\gamma^{\mu}\cdot\mbox{\boldmath $B$}_{u}(1,\vec 0)\ =\ \rlap/u
\end{equation*}
\item the fact that the Wigner rotation matrices, the $\chi$'s and $\gamma_o$
are all block-diagonal matrices, implying that they commute with the matrix
$\gamma_o$
which has scalar blocks.
\end{itemize}

\subsection{The heavy-mass limit}
Let us consider now the heavy-mass limit of \eqref{a}. By doing this,
we mean that we take $m_1,\,{m'}_1\longrightarrow+\infty$ and, at the same time,
we keep fixed the following ratios:
\begin{equation*}
v'\ =\ \frac{P'}{M'}\hspace{25mm}v\ =\ \frac{P}{M}
\end{equation*}
Therefore, we have:
\begin{align*}
&\dfrac{p_1}{m_1}\longrightarrow v & &\dfrac{{p'}_1}{{m'}_1}\longrightarrow v'\\
&u\longrightarrow v & &u'\longrightarrow v'\\
&\dfrac{k^o_1}{m_1}\longrightarrow 1 & &\dfrac{{k'}^o_1}{{m'}_1}
\longrightarrow 1\\
&k_2\longrightarrow \mbox{\boldmath $B$}^{-1}_{v}\,p_2	&
&k'_2\longrightarrow \mbox{\boldmath $B$}^{-1}_{v'}\,p_2\\
\end{align*}
As a consequence, since the scalar product of 4-vectors is invariant, we get:
\begin{align*}
(\mbox{\boldmath $B$}_v^{-1}p_2)^0 &= p_2.v &
(\mbox{\boldmath $B$}_{v'}^{-1}p_2)^0 &= p_2.v'
\end{align*}
We deduce from these formulae the invariance of $\norm{\overrightarrow{
\mbox{\boldmath $B$}_{v}^{-1}p_2}}^2$ and 
$\norm{\overrightarrow{\mbox{\boldmath $B$}_{v'}^{-1}p_2}}^2$:
\begin{equation}
\norm{\overrightarrow{\mbox{\boldmath $B$}_{v}^{-1}p_2}}^2 =
(p_2.v)^2 - m_2^2,\qquad
\norm{\overrightarrow{\mbox{\boldmath $B$}_{v'}^{-1}p_2}}^2=
(p_2.v')^2 - m_2^2\label{inv}
\end{equation}
Finally, the heavy mass limit of the relation \eqref{a} reads:
\begin{multline}\label{e}
_j\langle{^{2S+1}P_J}| O |{^1S_0}\rangle=\dfrac{1}{8}\,\dfrac{1}
{\sqrt{v_o\,v'_o}}\,
\int\,\frac{d\vec p_2}{(2\pi)^3}\,
\frac{1}{p^o_2}\,
\frac{\sqrt{(p_2.v')(p_2.v)}}{\sqrt{(p_2.v'+m_2)(p_2.v+m_2)}}\\
\times \mathrm { Tr}\left\{ {
O\,(1+\rlap/v)\,(m_2+\rlap/p_2)\,\left( {
\mbox{\boldmath $B$}_{v'}{\chi}^{\dagger}\mbox{\boldmath $B$}^{-1}_{v'}
}\right)\,(1+\rlap/v')
}\right\}\phi_j(\norm{\overrightarrow{\mbox{\boldmath $B$}_{v'}^{
-1}p_2}}^2)^{\displaystyle*}
\varphi(\norm{\overrightarrow{\mbox{\boldmath $B$}_{v}^{-1}p_2}}^2)
\end{multline}
\eqref{e} is the starting point for the calculation of the transition amplitudes
from a pseudoscalar state towards the $\left( {^{2S+1}P_J}\right)_j$ states
($D^{**}$ states):
that is the purpose of the next section. 

It is important here to stress that the expression (\ref{e}) is covariant. The
integration measure $d\vec p_2/p^o_2$ is invariant; so is the trace as obvious
from expressions (\ref{bchib}) for $\mbox{\boldmath $B$}_{v'}{\chi}^{\dagger}
\mbox{\boldmath $B$}^{-1}_{v'}$ and finally the arguments of the  wave functions
are invariant from (\ref{inv}). 

\section{Transition amplitudes and Isgur-Wise scaling}\label{sec-tau}
\subsection{Preliminary calculation}
Before calculating the spur which appears in \eqref{e}, we must evaluate
the $\mbox{\boldmath $B$}_{v'}{\chi}^{\dagger}\mbox{\boldmath $B$}^{-1}_{v'}$
terms, for each $\chi_{_{\left( {^{2S+1}P_J}\right)}}$. The procedure is quite
automatic:
\begin{enumerate}
\item We have to write each $\chi$, in the rest frame, in a covariant way,
that is using 4-vectors instead of 3-vectors. That is realized by introducing
the 4-vector $n^{\mu}=(1,\vec 0)$ and the expression of the Pauli matrices
in term of the Dirac matrices:
\begin{description}
\item{$\bullet\,({^{3}P_0})$:} Starting from $\chi_{_{\left( {^{3}P_0}\right)}
}$,
we get the $4\times 4$ matrix in the following way:
\begin{equation*}
\chi_{_{\left( {^{3}P_0}\right)}}^{\dagger}\ =\ -\dfrac{1}{\sqrt 3}\,
(\vec \sigma\cdot\vec k'_2)^{\dagger}
\ =\ -\dfrac{1}{\sqrt 3}\,(\vec \sigma\cdot\vec k'_2)
\ \leadsto\ -\dfrac{1}{\sqrt 3}\,{\underbrace{\gamma_5\gamma_0\,(\vec \gamma}_{
\bigl( \begin{smallmatrix}
\vec \sigma&0\\ 0&\vec \sigma
\end{smallmatrix} \bigr)}}
\cdot\vec k'_2)\,\frac{1+\gamma_0}{2}\ =\ \dfrac{1}{\sqrt 3}\,\gamma_5\,
(\vec \gamma
\cdot\vec k'_2)\,\frac{1+\gamma_0}{2}
\end{equation*} 
where the $\frac{1+\gamma_0}{2}$ factor in the right hand side gives the form
\begin{math}
\bigl( \begin{smallmatrix}
\chi&0\\ 0&0
\end{smallmatrix} \bigr)
\end{math} to the $\chi$. By introducing the $n^{\mu}$ mentioned above, this can
also be written as:
\begin{equation*}
\chi_{_{\left( {^{3}P_0}\right)}}^{\dagger}\ =\ \dfrac{1}{\sqrt 3}\,
\gamma_5\left[{-\rlap/{k'_2}\,+\,
{k'}^0_2\gamma_0}\right]\,\frac{1+\gamma_0}{2}
\ =\ \dfrac{1}{\sqrt 3}\,\gamma_5\left[{-\rlap/{k'_2}\,+\,{k'}^0_2
\gamma_{\mu}n^{\mu}}\right]\,\frac{1+\gamma_0}{2}
\end{equation*}
that is, finally,
\begin{equation*}
\chi_{_{\left( {^{3}P_0}\right)}}^{\dagger}\ =\ -\dfrac{1}{\sqrt 3}\gamma_5
\left[{\rlap/{k'_2}\,-\,{k'}^0_2\rlap/n}\right]\,\frac{1+\gamma_0}{2}
\end{equation*}
\item{$\bullet\,({^{1}P_1})$:} Introducing the 4-vector $e=(0,\vec e)$ (we
drop the
polarization index $m$ for the time being) such that $e\cdot n=0$, it is not
difficult to obtain:
\begin{equation*}
\chi_{_{\left( {^{1}P_1}\right)}}^{\dagger}\ =\ -(k'_2\cdot e
^{\displaystyle*})\,\frac{1+\gamma_0}{2}
\end{equation*}
\item{$\bullet\,({^{3}P_1})$:} Another manipulation is performed here in order
to deal
with the cross product, but the idea is still the same: to add a fourth
component.
\begin{equation*}
\left[ {\vec e\cdot(\vec k'_2\wedge\vec \sigma)}\right]^{\dagger}\ =\ 
\epsilon^{ijk}e^{\displaystyle*}_ik'_{2_j}\sigma_k\ \leadsto \ 
-\epsilon^{ijk}e^{\displaystyle*}_ik'_{2_j}\gamma_5\gamma_k\gamma_0\ \leadsto\ 
-\epsilon^{\mu\nu\rho\sigma}n_{\mu}e^{\displaystyle*}_{\nu}k'_{2_{\rho}}
\gamma_5\gamma_{\sigma}\gamma_0
\end{equation*}
Then, we get
\begin{equation*}
\chi_{_{\left( {^{3}P_1}\right)}}^{\dagger}\ =\ -\frac{i}{\sqrt 2}\,
\epsilon^{\mu\nu\rho\sigma}n_{\mu}e_{\nu}^{\displaystyle*}
k'_{2_{\rho}}\gamma_{\sigma}\gamma_5\,\frac{1+\gamma_0}{2}
\end{equation*}
\item{$\bullet\,({^{3}P_2})$:} In this last case, we generalize the
polarization tensor $e_{ij}$ into $e_{\mu\nu}$ which is
symmetrical and has a vanishing spur and obeys $n^{\mu}e_{\mu\nu}=0$. Thanks to
this last property, the $\chi$ writes:
\begin{equation*}
\chi_{_{\left( {^{3}P_2}\right)}}^{\dagger}\ =\ \gamma_5
\gamma^{\mu}{k'_2}^{\nu}e_{\mu\nu}^{\displaystyle*}\,\frac{1+\gamma_0}{2}
\end{equation*}
\end{description}
\item We now must evaluate the expressions $\mbox{\boldmath $B$}_{
v'}{\chi}^{\dagger}
\mbox{\boldmath $B$}^{-1}_{v'}$. That is achieved by inserting
the factor $\mbox{\boldmath $B$}_{v'}\mbox{\boldmath $B$}^{-1}_{v'}$
at the right places in the previous formulae
and extensively using \eqref{b}. Moreover, we introduce
the following notations:
\begin{align*}
\epsilon_{\mu\nu}&=\ \mbox{\boldmath $B$}_{v'}\,e_{\mu\nu} &
\epsilon_{\mu}&=\ \mbox{\boldmath $B$}_{v'}\,e_{\mu}
\end{align*}
As a result, the $\mbox{\boldmath $B$}_{v'}{\chi}^{\dagger}
\mbox{\boldmath $B$}^{-1}_{v'}$ read:
\begin{equation}\label{bchib}\left\{ \ {
\begin{split}
\mbox{\boldmath $B$}_{v'}
\chi_{_{{\left( {^{3}P_0}\right)}}}^{\dagger}\mbox{\boldmath $B$}^{-1}_{v'}\ &
=\ \dfrac{1}{\sqrt 3}
\left[{\rlap/{p_2}\,-\,(p_2\cdot v')\,\rlap/{v'}}\right]\gamma_5\,\frac{1+
\rlap/{v'}}{2}\\
\mbox{\boldmath $B$}_{v'}\chi_{_{{\left( {^{1}P_1}\right)}}}^{\dagger}
\mbox{\boldmath $B$}^{-1}_{v'}\ &=\ -(p_2\cdot \epsilon
^{\displaystyle*})\,\frac{1+\rlap/{v'}}{2}\\
\mbox{\boldmath $B$}_{v'}\chi_{_{{\left( {^{3}P_1}\right)}}}^{\dagger}
\mbox{\boldmath $B$}^{-1}_{v'}\ &=\ -\frac{i}{\sqrt 2}\,
\epsilon^{\mu\nu\rho\sigma}{v'}_{\mu}{\epsilon}_{\nu}^{\displaystyle*}
p_{2_{\rho}}\gamma_{\sigma}\gamma_5\,\frac{1+\rlap/{v'}}{2}\\
\mbox{\boldmath $B$}_{v'}
\chi_{_{{\left( {^{3}P_2}\right)}}}^{\dagger}\mbox{\boldmath $B$}^{-1}_{v'}\ 
&=\ -
\gamma^{\mu}{p_2}^{\nu}{\epsilon}_{\mu\nu}^{\displaystyle*}\gamma_5\,\frac{1+
\rlap/{v'}}{2}
\end{split} }\right.
\end{equation}
Notice the $\frac{1+\rlap/{v'}}{2}$ term in the previous relations; when
contracted
with the factor $(1+\rlap/{v'})$ of the spur in \eqref{e}, the result will be
$(1+\rlap/{v'})$
since $\rlap/{v'}^2=1$. So, from now on, we will drop it from the expressions
$\mbox{\boldmath $B$}_
{v'}{\chi}^{\dagger}\mbox{\boldmath $B$}^{-1}_{v'}$.
\end{enumerate}

\subsection{The transition amplitudes}
\subsubsection{Definitions}
On inserting the expressions \eqref{bchib} from the previous section into
\eqref{e}, three different kinds of integrals appear using the covariance of
the matrix elements, \eqref{e}. Therefore, we may introduce the following
definitions regarding the result of these integrals:
\begin{eqnarray*}
A^{(j)}&=&
\int\,\frac{d\vec p_2}{(2\pi)^3}\,
\frac{1}{p^o_2}\,
\frac{\sqrt{(p_2\cdot v')(p_2\cdot v)}}{\sqrt{(p_2\cdot v'+m_2)(p_2\cdot v+m_2)
}}
\phi_j(\norm{\overrightarrow{\mbox{\boldmath $B$}_{v'}^{-1}p_2}}^2)^{
\displaystyle*}
\varphi(\norm{\overrightarrow{\mbox{\boldmath $B$}_{v}^{-1}p_2}}^2)\\
B^{(j)}\,v^{\mu}\ +\ {B'}^{(j)}\,{v'}^{\mu}&=&
\int\,\frac{d\vec p_2}{(2\pi)^3}\,
\frac{1}{p^o_2}\,
\frac{\sqrt{(p_2\cdot v')(p_2\cdot v)}}{\sqrt{(p_2\cdot v'+m_2)(p_2\cdot v+m_2)
}}
\phi_j(\norm{\overrightarrow{\mbox{\boldmath $B$}_{v'}^{-1}p_2}}^2)^{
\displaystyle*}
\varphi(\norm{\overrightarrow{\mbox{\boldmath $B$}_{v}^{-1}p_2}}^2)\,p_2^{\mu}\\
D_1^{(j)}(v^{\mu}\,{v'}^{\nu}+v^{\nu}\,{v'}^{\mu})&+&D_2^{(j)}\,v^{\mu}\,v^{\nu}
\ +\ D_3^{(j)}\,{v'}^{\mu}\,{v'}^{\nu}\ +\ D_4^{(j)}\,g^{\mu\nu}\\
&=&\int\,\frac{d\vec p_2}{(2\pi)^3}\,
\frac{1}{p^o_2}\,
\frac{\sqrt{(p_2\cdot v')(p_2\cdot v)}}{\sqrt{(p_2\cdot v'+m_2)(p_2\cdot v+m_2)
}}
\phi_j(\norm{\overrightarrow{\mbox{\boldmath $B$}_{v'}^{-1}p_2}}^2)^{
\displaystyle*}
\varphi(\norm{\overrightarrow{\mbox{\boldmath $B$}_{v}^{-1}p_2}}^2)\,p_2^{\mu}
\,p_2^{\nu}\\
\end{eqnarray*}
where the $A$, $B$'s and the $D$'s are function of $w\,=\,v\cdot v'$.

\subsubsection{Results}
We now have all the pieces to reduce \eqref{e} further for all the
states $\left( {jJ^P}\right)$. Straightforward calculations and a little
diracology lead to:
\begin{eqnarray*}
\langle{{\scriptstyle\frac{1}{2}}\,0^+}| O |{^1S_0}\rangle&=&
\dfrac{1}{8\sqrt 3}\,\dfrac{1}{\sqrt{v_o\,{v'}_o}}\,
\mathrm {Tr}\left\{ {O\,(1+\rlap/v)\,(1-\rlap/v')\,\gamma_5
}\right\} \\
&&\times\left[ {(1+v\cdot v')\left( {m_2\,B^{(1/2)}+D_1^{(1/2)}+D_2^{(1/2)}}
\right)
\ +\ 3\,D_4^{(1/2)}}\right]\\ \\ \\
\langle{{\scriptstyle\frac{1}{2}}\,1^+}| O |{^1S_0}\rangle&=&
\dfrac{1}{8\sqrt 3}\,\dfrac{1}{\sqrt{v_o\,{v'}_o}}\,
\left( {m_2\,B^{(1/2)}+D_1^{(1/2)}+D_2^{(1/2)}}\right)\\
&&\times\left[ {\mathrm {Tr}\left\{ {O\,(1+\rlap/v)\,(1+\rlap/v')}\right\}\,
(v\cdot\epsilon^{\displaystyle*})\ +\ i\,\epsilon^{\mu\nu\rho\sigma}\,{v'}_{
\mu}\,
\epsilon^{\displaystyle*}_{\nu}\,v_{\rho}\,\mathrm {Tr}\left\{ {O\,(1+\rlap/v)\,
\gamma_{\sigma}\,(1-\rlap/v')\,\gamma_5}\right\}
}\right]\\
&+&
\dfrac{1}{8\sqrt 3}\,\dfrac{1}{\sqrt{v_o\,{v'}_o}}\,
D_4^{(1/2)}\\
&&\times\left[ {\mathrm {Tr}\left\{ {O\,(1+\rlap/v)\,\rlap/{\epsilon^{
\displaystyle*}}\,
(1+\rlap/v')}\right\}
\ +\ i\,\epsilon^{\mu\nu\rho\sigma}\,{v'}_{\mu}\,
\epsilon^{\displaystyle*}_{\nu}\,\mathrm {Tr}\left\{ {O\,(1+\rlap/v)\,\gamma_{
\rho}\,
\gamma_{\sigma}\,(1-\rlap/v')\,\gamma_5}\right\}
}\right]\\ \\ \\
\langle{{\scriptstyle\frac{3}{2}}\,1^+}| O |{^1S_0}\rangle&=&
\dfrac{1}{4\sqrt 6}\,\dfrac{1}{\sqrt{v_o\,{v'}_o}}\,
\left( {m_2\,B^{(3/2)}+D_1^{(3/2)}+D_2^{(3/2)}}\right)\\
&&\times\left[ {-\mathrm {Tr}\left\{ {O\,(1+\rlap/v)\,(1+\rlap/v')}\right\}\,
(v\cdot\epsilon^{\displaystyle*})\ +\ \frac{i}{2}\,\epsilon^{\mu\nu\rho\sigma}
\,{v'}_{\mu}\,
\epsilon^{\displaystyle*}_{\nu}\,v_{\rho}\,\mathrm {Tr}\left\{ {O\,(1+\rlap/v)\,
\gamma_{\sigma}\,(1-\rlap/v')\,\gamma_5}\right\}
}\right]\\
&+&
\dfrac{1}{4\sqrt 6}\,\dfrac{1}{\sqrt{v_o\,{v'}_o}}\,
D_4^{(3/2)}\\
&&\times\left[ {-\mathrm {Tr}\left\{ {O\,(1+\rlap/v)\,\rlap/{\epsilon^{
\displaystyle*}}\,
(1+\rlap/v')}\right\}
\ +\ \frac{i}{2}\,\epsilon^{\mu\nu\rho\sigma}\,{v'}_{\mu}\,
\epsilon^{\displaystyle*}_{\nu}\,\mathrm {Tr}\left\{ {O\,(1+\rlap/v)\,\gamma_{
\rho}\,
\gamma_{\sigma}\,(1-\rlap/v')\,\gamma_5}\right\}
}\right]\\ \\ \\
\langle{{\scriptstyle\frac{3}{2}}\,2^+}| O |{^1S_0}\rangle&=&
-\dfrac{1}{8\sqrt 3}\,\dfrac{1}{\sqrt{v_o\,{v'}_o}}\,
\left( {m_2\,B^{(3/2)}+D_1^{(3/2)}+D_2^{(3/2)}}\right)\,
\mathrm {Tr}\left\{ {O\,(1+\rlap/v)\,\gamma^{\mu}\,(1-\rlap/v')
\,\gamma_5}\right\}\,
v^{\nu}\,\epsilon^{\displaystyle*}_{\mu\nu}\\
&-&
\dfrac{1}{8\sqrt 3}\,\dfrac{1}{\sqrt{v_o\,{v'}_o}}\,
D_4^{(3/2)}\,\mathrm {Tr}\left\{ {O\,(1+\rlap/v)\,\gamma^{\mu}\,\gamma^{\nu}\,
(1-\rlap/v')\,\gamma_5}\right\}\,
\epsilon^{\displaystyle*}_{\mu\nu}
\end{eqnarray*}

\subsubsection{Vector and axial matrix elements}
Up to now, we did not take into account an explicit form for the current
operator $O$. Let us see what becomes of the preceding relations when we
use $O=V_{\lambda}=\gamma_{\lambda}$ and $O=A_{\lambda}=\gamma_{\lambda}
\gamma_5$.
\begin{multline*}
\langle{{\scriptstyle\frac{1}{2}}\,0^+}|\,V_{\lambda}\,|{^1S_0}\rangle=0\\
\shoveleft{\langle{{\scriptstyle\frac{1}{2}}\,0^+}|\,A_{\lambda}\,|{^1S_0}
\rangle=
-\frac{1}{2\sqrt 3}\,\frac{1}{\sqrt{v_o\,v_o'}}\,(v_{\lambda}\,-\,v_{\lambda}')
\left\{ {(1+v\cdot v')\left( {m_2\,B^{(1/2)}+D_1^{(1/2)}+D_2^{(1/2)}}\right)
\ +\ 3\,D_4^{(1/2)}}\right\}}\\
\shoveleft{\langle{{\scriptstyle\frac{1}{2}}\,1^+}|\,V_{\lambda}\,|{^1S_0}
\rangle=
\frac{1}{\sqrt{12}}\,\frac{1}{\sqrt{v_o\,v_o'}}\,\left[ { (v\cdot \epsilon^{
\displaystyle*})
v_{\lambda}'\ +\ (1-v\cdot v')\epsilon^{\displaystyle*}_{\lambda}}\right]}\\
\shoveright{\times\left\{ {(1+v\cdot v')\left( {m_2\,B^{(1/2)}+D_1^{(1/2)}+
D_2^{(1/2)}}\right)
\ +\ 3\,D_4^{(1/2)}}\right\}}\\
\shoveleft{\langle{{\scriptstyle\frac{1}{2}}\,1^+}|\,A_{\lambda}\,|{^1S_0}
\rangle=
-\frac{i}{\sqrt{12}}\,\frac{1}{\sqrt{v_o\,v_o'}}\,\epsilon_{\lambda\sigma\rho
\tau}\,
v^{\sigma}\,\epsilon^{{\displaystyle*}\rho}\,{v'}^{\tau}
\left\{ {(1+v\cdot v')\left( {m_2\,B^{(1/2)}+D_1^{(1/2)}+D_2^{(1/2)}}\right)
\ +\ 3\,D_4^{(1/2)}}\right\}}\\
\end{multline*}
for the $j=1/2$ multiplet, and
\begin{multline*}
\langle{{\scriptstyle\frac{3}{2}}\,1^+}|\,V_{\lambda}\,|{^1S_0}\rangle=
\frac{1}{2\sqrt 6}\,\frac{1}{\sqrt{v_o\,v_o'}}\,\left[ { (v\cdot \epsilon^{
\displaystyle*})
\left\{ { v_{\lambda}'(v\cdot v'-2)-3\,v_{\lambda} } \right\}\ +\ 
\epsilon^{\displaystyle*}_{\lambda}(1-v\cdot v')(1+v\cdot v')}\right]\\
\shoveright{\times\left\{ {m_2\,B^{(3/2)}+D_1^{(3/2)}+D_2^{(3/2)} }\right\}
}\\
\shoveleft{\langle{{\scriptstyle\frac{3}{2}}\,1^+}|\,A_{\lambda}\,|{^1S_0}
\rangle=
-\frac{i}{2\sqrt 6}\,\frac{1}{\sqrt{v_o\,v_o'}}\,(1+v\cdot v')\,\epsilon_{
\lambda\sigma\rho\tau}\,
v^{\sigma}\,\epsilon^{{\displaystyle*}\rho}\,{v'}^{\tau}
\left\{ {m_2\,B^{(3/2)}+D_1^{(3/2)}+D_2^{(3/2)} }\right\}
}\\
\shoveleft{\langle{{\scriptstyle\frac{3}{2}}\,2^+}|\,V_{\lambda}\,|{^1S_0}
\rangle=
\frac{i}{2}\,\frac{1}{\sqrt{v_o\,v_o'}}\,\epsilon_{\lambda\alpha\mu\beta}\,
\epsilon^{{\displaystyle*}\mu\nu}\,v_{\nu}\,v^{\alpha}\,{v'}^{\beta}
\left\{ {m_2\,B^{(3/2)}+D_1^{(3/2)}+D_2^{(3/2)} }\right\}
}\\
\shoveleft{\langle{{\scriptstyle\frac{3}{2}}\,2^+}|\,A_{\lambda}\,|{^1S_0}
\rangle=
\frac{1}{2}\,\frac{1}{\sqrt{v_o\,v_o'}}\,\left[ {
(1+v\cdot v')\,\epsilon^{\displaystyle*}_{\lambda\nu}\,v^{\nu}\ -\ v^{\mu}\,v^{
\nu}\,
\epsilon^{\displaystyle*}_{\mu\nu}\,{v'}_{\lambda}
}\right]
\left\{ {m_2\,B^{(3/2)}+D_1^{(3/2)}+D_2^{(3/2)} }\right\}
}\\
\end{multline*}
for the $j=3/2$ multiplet. As expected, all these matrix elements are
manifestly covariant.

\subsection{Isgur-Wise scaling}
Looking at the results of the last section, it is fairly obvious that all
the matrix elements can be written using two differents quantities,
$\tau_{1/2}(v\cdot v')$
and $\tau_{3/2}(v\cdot v')$, following the notation introduced by Isgur
and Wise in \cite{IW}.
These $\tau_j(v\cdot v')$'s are:
\begin{eqnarray}\label{tau}
\tau_{_{1/2}}(v\cdot v')&=&\frac{1}{2\sqrt 3}\,\left\{ {
(1+v\cdot v')\left( {m_2\,B^{(1/2)}+D_1^{(1/2)}+D_2^{(1/2)}}\right)
\ +\ 3\,D_4^{(1/2)}}\right\}\nonumber\\
\tau_{_{3/2}}(v\cdot v')&=&\frac{1}{\sqrt 3}\,\left\{ {m_2\,B^{(3/2)}+
D_1^{(3/2)}+D_2^{(3/2)} }\right\}
\end{eqnarray}
Two remarks though: our definition of the state
${|{\scriptstyle\frac{1}{2}}\,1^+\rangle}$
differs from the one given in \cite{IW} by an overall minus sign; and,
secondly, we are using
another normalization of the states. As a consequence, there is an overall
factor of $
2\sqrt{m_P\,m_{D^{**}}}\sqrt{v_o\,v_o'}$
between our transition amplitudes and the transition amplitudes written
in \cite{IW}.

\section{The Bjorken sum rule}\label{sec-bj}
Generally speaking, the Bjorken sum rule is a way of expressing duality
(between quarks
and hadrons), which, in the case of the heavy quark mass limit, is an exact
duality. Mathematically, this rule reads \cite{alain}:
\begin{multline}\label{gsum}
 \bar{h}_{\mu \nu} (\vec{v} , \vec{v}\,{'}) \equiv {\left( {
\sum_n <n, \vec{P}'|O^{\mu}|0, \vec{P}>
\times (<n,\vec P'|O^{\nu}|0, \vec{P}>)^*}\right) }_{\displaystyle{
\text{where }m_1\to\infty}}\\
= \frac12 \sum_{s_1, s'_1} \left [ \bar{u}_{s'_1}(\vec{v}\,') O^{\mu} u_{s_1}
(\vec{v}) \right ] \left [ \bar{u}_{s'_1} (\vec{v}\,') O^{\mu} u_{s_1}(\vec{v})
\right ]^* 
\equiv \bar{h}_{\mu \nu}^{\text{free quark}} 
\end{multline}
where $\bar{h}_{\mu \nu} (\vec{v} , \vec{v}\,')$ is the ``hadronic tensor''
which describes
the transition between an initial meson state to all possible meson states
containing
a heavy quark, and where
$\bar{h}_{\mu \nu}^{\text{free quark}}$ is the analog for free quarks. Note that
$n$ is used to label the vectors of the {\em complete} basis
$\{|n, \vec{P}>\}$.\par\vspace{5mm}
By using the expressions of the current matrix elements which are involved in
the
definition of $\bar{h}_{\mu \nu} (\vec{v} , \vec{v}\,')$ and expanding around
$w=\vec{v}\cdot\vec{v}\,'=1$, we get the new following Bjorken sum rule:
\begin{equation}\label{sum}
\sum_{n}\left( {
\abs{\tau^{(n)}_{_{1/2}}(1)}^2\ +\ 2\,\abs{\tau^{(n)}_{_{3/2}}(1)}^2
} \right)\ =\ \rho^2\ -\ \dfrac14
\end{equation}
where $\rho^2$ is the slope of the elastic ground state Isgur-Wise scaling
function and
where the superscripts $(n)$ characterize the radial excitations of the
P wave states, to which the results of the preceding section obviously apply
(these $(n)$ correspond exactly to the labels $n$
used in \eqref{gsum}).
In \cite{alain}, the physical meaning and a detailed demonstration of that
sum rule
have been given. In the following, we are just going to verify that
\eqref{sum} still
holds with the covariant formalism.
\subsection{The ${\boldsymbol {\tau_j(1)}}$'s}
Checking \eqref{sum} begins with evaluating the $\tau_j$'s at $v\cdot v'=1$.
Since
they are given in a covariant way, let's choose the particular frame of
reference where:
\begin{eqnarray*}
\left|
\begin{array}{rcll}
\vec {v'}&=&\vec 0 &\text{(then }v'_0=1)\\
\vec v&\neq&\vec 0\ \ \text{ with } \vec v \text{ small}  \quad&(
\text{then }v_0\simeq 1+\dfrac12\,{\vec v}^2)\\
\end{array}\right.
\end{eqnarray*}
Therefore, to the second order in $\vec v$, we get:
\begin{eqnarray*}
p\cdot v'&=&p_0\\
p\cdot v&\simeq&p_0\ -\ \vec p\cdot\vec v\ +\ \dfrac12\,p_0\,{\vec v}^2\\
v\cdot v'&\simeq&1\ +\ \dfrac12\,{\vec v}^2
\end{eqnarray*}
and also
\begin{eqnarray*}
\norm{\overrightarrow{\mbox{\boldmath $B$}_{v'}^{-1}p}}^2&=&{\vec p}\,^2\\
\norm{\overrightarrow{\mbox{\boldmath $B$}_{v}^{-1}p}}^2&
\simeq&{\vec p}\,^2\ -\ 2\,p_0\,\vec p\cdot\vec v
\end{eqnarray*}
Then, we substitute these expressions into \eqref{tau} and expand to the
first order 
in $\vec v$ the
$\phi$ functions of \eqref{tau}. Owing to rotational invariance, the terms of
the form
$\vec v/{\vec v}^2$ vanish and, when
we take $\vec v$ equal to zero, we are left  with:
\begin{equation}\label{t1}
\tau^{(n)}_j(1)=\int\,\dfrac{p^2\,dp}{(2\pi)^2}\,\phi_j^{{
\displaystyle *}\,(n)}(p^2)\,F_j(p^2)
\end{equation}
with $p=\norm{\vec p}$ and where
\begin{eqnarray*}
F_{_{1/2}}(p^2)&=&-\dfrac{1}{3\sqrt 3}\left\{ {
\varphi(p^2)\,\frac{p^2}{m+p_0}\,\left( {3+\frac{m}{p_0}}\right)\ +\ 4\,
\frac{d\varphi}{d p^2}(p^2)\,p_0\,p^2 } \right\}\\
F_{_{3/2}}(p^2)&=&-\dfrac{1}{3\sqrt 3}\left\{ {
\varphi(p^2)\,\frac{p^2}{m+p_0}\,\dfrac{m}{p_0}\ +\ 4\,
\frac{d\varphi}{d p^2}(p^2)\,p_0\,p^2 } \right\}\\
\end{eqnarray*}

\subsection{Connection between the ${\boldsymbol {F_j}}$'s and the sum
of the ${\boldsymbol {\tau_j}}$'s}
Let us work in the part ${\cal H}_{\vec j}$
of the total Hilbert space ${\cal H}$ describing the wave functions, that is,
let us
fix the spin of the heavy quark (which we know is irrelevant). Then, the
partial wave function
writes:
\begin{equation*}
\Psi^{(n)\,\mu}_j(\vec p)\ =\ \sum_m\Phi^{(n)\,m}_j(\vec p)\langle\,1\ m\  
{1/2}\ \mu-m\ |\ j\ \mu\,\rangle\,\chi^{\mu-m}
\end{equation*}
In ${\cal H}_{\vec j}$, the {\em closure relation} reads:
\begin{equation}\label{ferm}
\sum_{j,n,\mu}{\Psi^{(n)\,\mu}_j}^{\dagger}(\vec p)\Psi^{(n)\,\mu}_j(\vec p\,')
\ =\ (2\pi)^3\,\delta(\vec p - \vec p\,')
\end{equation}
If we multiply \eqref{ferm} by
$\int d\Omega\,d\Omega'\,Y^{m_1}_1(\Omega)\,Y^{m'_1}_1(\Omega')$ and sandwich
the result
between $\chi^{\nu}$ on the left and ${\chi^{\nu'}}^{\dagger}$ on the right,
we obtain:
\begin{equation*}
\sum_{j,n}\frac{4\pi}{3}\,pp'\,\phi_j^{{\displaystyle *}\,(n)}(p^2)\,
\phi_j^{(n)}({p'}^2)
\,\delta_{\mu\mu'}\,\langle\,1\ m_1\  
{1/2}\ \nu\ |\ j\ \mu\,\rangle\,\langle\,1\ m'_1\  
{1/2}\ \nu'\ |\ j\ \mu'\,\rangle\ =\ (2\pi)^3\,\frac{\delta(p-p')}{{p'}^2}\,
\delta_{m_1\,m'_1}\,
\delta_{\nu\,\nu'}
\end{equation*}
where $p=\norm{\vec p}$ and $p'=\norm{\vec p\,'}$.
Finally, by multiplying this last relation by $\langle\,1\ m_1\  
{1/2}\ \nu\ |\ J\ M\,\rangle\,\langle\,1\ m'_1\  
{1/2}\ \nu'\ |\ J'\ M'\,\rangle$ and summing the result over
$m_1,\,m'_1,\,\nu,\,\nu'$, we get:
\begin{equation*}
\sum_n\phi_j^{{\displaystyle *}\,(n)}(p^2)\,\phi_j^{(n)}({p'}^2)\ =\ 6\pi^2\,
\frac{\delta(p-p')}
{p{p'}^3}
\end{equation*}
for each value of $j$. We now have all the pieces to calculate $\sum \abs{
\tau_j}^2$; starting
from \eqref{t1} and summing its squared norm, we get:
\begin{equation*}
\sum_n\abs{\tau_j^{(n)}(1)}^2\ =\ \dfrac{3}{8\pi^2}\,\int\,dp\,\abs{F_j(p^2)}^2
\end{equation*}

\subsection{Checking the Bjorken sum rule for ${\boldsymbol {\rho^2}}$}
We now are able to evaluate the left hand side of \eqref{sum}. After integrating
by parts (the integral of the divergence vanishes), we get:
\begin{equation*}
\sum_{n}
\abs{\tau^{(n)}_{_{1/2}}(1)}^2\ +\ 2\,\sum_{n}\abs{\tau^{(n)}_{_{3/2}}(1)}^2\ 
=\ 
\dfrac23\,\int\frac{dp}{(2\pi)^2}\,\left\{ {\varphi(p^2)\,\varphi(p^2)^{
\displaystyle *}
\left[{\dfrac14\,\frac{p^4}{p_0^2}\ -\ \frac{p_0+2m}{p_0+m}\,p^2
}\right]\ +\ \frac{d\varphi}{d p^2}(p^2)\,\frac{d\varphi}{d p^2}(p^2)^{
\displaystyle *}\,4\,p_0^2\,p^4
}\right\}
\end{equation*}
Regarding the right hand side of \eqref{sum}, we start from the expression of
the slope
$\rho^2$ given by the equation $(29)$ of \cite{raynal}:
\begin{equation*}
\rho^2\  = \  \frac13  \int \frac{d\vec p}{(2\pi)^3}\, 
[\vec\nabla\,p^0\varphi(\vec p)]^{\displaystyle*}.[\vec\nabla\,p^0\varphi(
\vec p)]
\ +\  \int  \frac{d\vec p}{(2\pi)^3}\, [\frac23+\frac14\,\frac{m^2}{(p^0)^2}-
\frac13\,
\frac{m}{p^0{+}m}]\;\varphi(\vec p)^{\displaystyle*}\varphi(\vec p)
\end{equation*}
 Then, we use 
\begin{equation*}
\vec\nabla(p_0\varphi)\ =\ p_0\,\vec\nabla\varphi\ +\ \varphi\,\vec\nabla p_0
\ =\ 2\,p_0\,\vec p\,\frac{d\varphi}{d p^2}\ +\ \dfrac{\vec p}{p_0}\,\varphi
\end{equation*}
and, after another integration by parts, we get
\begin{equation*}
\rho^2\ -\ \dfrac14\ =\ 
\dfrac23\,\int\frac{dp}{(2\pi)^2}\,\left\{ {\varphi(p^2)\,\varphi(p^2)^{
\displaystyle *}
\left[{\dfrac14\,\frac{p^4}{p_0^2}\ -\ \frac{p_0+2m}{p_0+m}\,p^2
}\right]\ +\ \frac{d\varphi}{d p^2}(p^2)\,\frac{d\varphi}{d p^2}(p^2)^{
\displaystyle *}\,4\,p_0^2\,p^4
}\right\}
\end{equation*}
showing that \eqref{sum} is indeed verified.

\section{Conclusion}\label{sec-con}
In this paper, 
we have considered the $B\to D^{\ast\ast}$ type transitions quark models \`a
la BT
in the heavy quark limit. As already shown for the $B\to D^{(\ast)}$
transitions \cite{raynal}, these models verify
covariance and heavy quark symmetry. Consequently, all the hadronic matrix
elements between a pseudoscalar
state and a P-wave meson state, both containing a quark with infinite mass,
can be expressed 
using two different functions, namely $\tau_{\scriptscriptstyle{1/2}}$ and
$\tau_{\scriptscriptstyle{3/2}}$, which we have computed in these models:
\begin{multline*}
\tau_{\scriptscriptstyle{1/2}}(w)\ =\ \dfrac{1}{2\sqrt 3}\,\int\,\frac{d
\vec p_2}{(2\pi)^3}\,\frac{1}{p^o_2}\,
\frac{\sqrt{(p_2\cdot v')(p_2\cdot v)}}{\sqrt{(p_2\cdot v'+m_2)(p_2\cdot v+m_2)
}}
\phi_{\frac12}((p_2.v')^2 - m_2^2)^{\displaystyle*}
\varphi((p_2.v)^2 - m_2^2)\\
\times\,\frac{(p_2\cdot v)(p_2\cdot {v'}+m_2)\ -\ (p_2\cdot {v'})(p_2\cdot {v'}
+(v\cdot {v'})m_2)
\ +\ (1-v\cdot {v'})m_2^2}{1-v\cdot {v'}}
\end{multline*}
and
\begin{multline*}
\tau_{\scriptscriptstyle{3/2}}(w)\ =\ 
\dfrac{1}{\sqrt 3}\,\dfrac{1}{1-(v\cdot {v'})^2}\,
\int\,\frac{d\vec p_2}{(2\pi)^3}\,\frac{1}{p^o_2}\,
\frac{\sqrt{(p_2\cdot v')(p_2\cdot v)}}{\sqrt{(p_2\cdot v'+m_2)(p_2\cdot v+m_2)
}}
\phi_{\frac32}((p_2.v')^2 - m_2^2)^{\displaystyle*}
\varphi((p_2.v)^2 - m_2^2)\\
\times\ \left\{ {\dfrac32\,\frac{1}{1+v\cdot {v'}}\,\left[ {\,p_2\cdot (v+{v'})
}\right]^2\ -\ 
(p_2\cdot v)(2\,p_2\cdot {v'}-m_2)\ -\ (p_2\cdot {v'})\left[  {\,p_2\cdot {v'}
+(v\cdot {v'})m_2}\right]\ -\ \dfrac{1-v\cdot {v'}}{2}\,m_2^2
}\right\}
\end{multline*}
In the two last equations the apparent poles at $v\cdot v'=1$ are canceled
by zeros in the numerators.

We have also checked the validity of the Bjorken sum rule already demonstrated
in \cite{alain}.
These results are general for all quark models \`a la BT, independently of the
precise dynamics.
To complete the calculation, it is necessary to have an explicit form for the
radial functions
$\phi_j$ and $\varphi$. Therefore, what remains to be done is to take a
hamiltonian and
then use it to solve a rest-frame Schr\"odinger-type equation, leading to
wave functions from
which the radial parts can be extracted. That will be done in a forthcoming
paper.
Of course, all the difficulties lie in the choice of the hamiltonian
describing the meson
(as an example, the resulting values of $\rho^2$ must not be too
high\footnote{The experimental value of $\hat
\rho^2=0.87 \pm 0.12 \pm 0.08$ from CLEO II is not easy to compare to $\rho^2$
in our model since $1/m_1$ corrections and radiative corrections should
be considered. Still we would like to avoid a too large discrepancy.}).
When this is achieved, we will be able to compute, consistently with the
Bjorken sum rule, $\rho^2$ and
the rate of the semileptonic decay $B\to D^{\ast\ast}\,l\nu$.


\begin{thebibliography}{99}
\bibitem{raynal}A. Le Yaouanc, L. Oliver, O. P\`ene and J.-C. Raynal, 
Phys. Lett. B {\bf 365}, 319 (1996).
\bibitem{BT} B. Bakamjian and L. H. Thomas, Phys. Rev. {\bf 92}, 1300
(1953).
\bibitem{alain} A. Le Yaouanc, L. Oliver, O. P\`ene et J.-C. Raynal,
LPTHE-Orsay 96/05.
\bibitem{bjorken} J. D. Bjorken,
talk at Les Rencontres de Physique de la Vall\'ee d'Aoste, La Thuile, Italy,
SLAC Report
SLAC-PUB-5278, 1990 (unpublished); Slac 1990 summer institute, SLAC Report-378,
pp. 167-198 ; J. D. Bjorken, I. Dunietz and J. Taron, Nucl. Phys.
{\bf B371}, (1992). 
\bibitem{IW} N. Isgur and M. B. Wise, Phys. Rev. {\bf D43}, 819 (1991).

\bibitem{close} F. E. Close and A. Wambach, Nucl. Phys.  {\bf B412},
169 (1994).
\bibitem{wambach} A. Wambach, Nucl. Phys. {\bf B434} (1995) 647. 



\end{thebibliography}
\end{document}